# *Deep Learning-based Accelerated MR Cholangiopancreatography without Fully-sampled Data*

*Jinho Kim[1,2], Marcel Dominik Nickel[2], and Florian Knoll[1,3]*

*[1]Department of Artificial Intelligence in Biomedical Engineering, Friedrich-Alexander-Universität Erlangen-Nürnberg, Germany, [2] Research & Clinical Translation, Magnetic Resonance, Siemens Healthineers, Erlangen, Germany, [3]Center for Advanced Imaging Innovation and Research (CAI²R), Department of Radiology, New York University Grossman School of Medicine, New York, NY, USA*

Correspondence to: Jinho Kim
Computational Imaging Lab
Department of Artificial Intelligence in Biomedical Engineering
Friedrich-Alexander-Universität Erlangen-Nürnberg
Werner-Von-Siemens-Straße. 61, D-91054 Erlangen, Germany
E-mail: jinho.kim@fau.de

Number of words (abstract): 210
Number of words (body): ca. 3,764
Number of figures: 4
Number of tables: 2

# Abstract

The purpose of this study was to accelerate MR cholangiopancreatography (MRCP) acquisitions using deep learning-based (DL) reconstruction at 3T and 0.55T. A total of 35 healthy volunteers underwent conventional two-fold accelerated MRCP scans at field strengths of 3T and 0.55T. We trained DL reconstructions using two different training strategies, supervised (SV) and self-supervised (SSV), with retrospectively six-fold undersampled data obtained at 3T. We then evaluated the DL reconstructions against standard techniques, parallel imaging (PI) and compressed sensing (CS), focusing on peak signal-to-noise ratio (PSNR) and structural similarity (SSIM) as metrics. We also tested DL reconstructions with prospectively accelerated acquisitions and evaluated their robustness when changing fields strengths from 3T to 0.55T. DL reconstructions demonstrated a reduction in average acquisition time from 599/542 to 255/180 seconds for MRCP at 3T/0.55T. In both retrospective and prospective undersampling, PSNR and SSIM of DL reconstructions were higher than those of PI and CS. At the same time, DL reconstructions preserved the image quality of undersampled data, including sharpness and the visibility of hepatobiliary ducts. In addition, both DL approaches produced high-quality reconstructions at 0.55T. In summary, DL reconstructions trained for highly accelerated MRCP enabled a reduction in acquisition time by a factor of 2.4/3.0 at 3T/0.55T while maintaining the image quality of conventional acquisitions.

# Introduction

Magnetic resonance cholangiopancreatography (MRCP) is a non-invasive imaging technique used for diagnosing diseases of the hepatobiliary system, providing detailed views of the ductal structures and related pathologies[1–4]. Initially performed using 2D thick-slice acquisitions, MRCP has evolved into 3D imaging, improving image quality and providing comprehensive multidimensional views. However, these improvements have resulted in longer acquisition times, which are prone to motion artifacts[5]. There are two acquisition methods to address these challenges: breath-hold and triggered free-breathing. Due to the limitations of breath-holding, the triggered free-breathing acquisition method is more practical in clinical applications.

Triggered techniques allow patients to breathe naturally during the scan while minimizing motion artifacts caused by breathing. The prospective acquisition correction (PACE) technique has significantly enhanced the quality of triggered free-breathing 3D MRCP. Compared to conventional respiratory-based triggering methods, PACE triggering produces fewer motion artifacts and provides sharper anatomical contours[6]. Furthermore, Asbach et al.[7] highlighted that PACE-triggered free-breathing 3D MRCP (PACE-MRCP) significantly improves the visualization of hepatobiliary and pancreatic ductal structures compared to breath-holding 3D MRCP, providing more detailed images and greater patient comfort during the scan. Despite these advantages, PACE-MRCP may require long acquisition times due to irregular breathing patterns.

To address the challenge of long acquisition times in PACE-MRCP, recent studies have focused on accelerating $k$-space data acquisition[8]. This acceleration is typically achieved by undersampling the $k$-space data, which reduces acquisition time but increases the risk of aliasing artifacts. Various reconstruction methods, such as parallel imaging (PI) and compressed sensing (CS), have been developed to address these artifacts[9,10]. Given uncorrelated coil sensitivity profiles, PI significantly reduces scan time by exploiting correlations between multiple coil sensitivity profiles[11]. CS leverages the incoherence of undersampling patterns, leading to uncorrelated aliasing artifacts, and enables image reconstruction from highly undersampled and sparse $k$-space data[12].

In recent years, deep learning-based (DL) methods have gained popularity in MR reconstruction[13], particularly in physical model-driven DL approaches. These methods are inspired by traditional iterative optimization problems and use unrolled network architectures. Model parameters are typically determined through supervised training for prospective use on unseen data. The overall models can be viewed as generalized functions mapping undersampled data to high quality reconstructions. These methods have shown remarkable performance for the reconstruction of highly accelerated MRI[14,15].

Aggarwal et al.[14] introduced a supervised DL reconstruction method (SV) that requires fully sampled ground truth, but obtaining such data, particularly for MRCP, is challenging because it is not possible to acquire fully sampled ground truth data. Given the common use of accelerated MRCP in clinical settings, it might be possible to train SV using moderately accelerated MRCP acquisitions as substitutes for fully sampled data. Meanwhile, to address the lack of fully sampled data, Yamman et al.[15] proposed a self-supervised DL reconstruction method (SSV) and has since been adapted to various applications[16–18]. One goal of our study was to evaluate whether synthesized ground truth data obtained with conventional parallel imaging in combination with supervised training, or self-supervised training is the better choice, for accelerated MRCP.

Another goal of our study was to evaluate the generalization of DL reconstruction model towards changes in magnet field strengths. Going to a different field strength changes the signal-to-noise ratio (SNR) and it has been demonstrated that DL reconstructions can be sensitive to changes in SNR[19]. We therefore assessed the generalizability of our DL models by training them on high-field MRCP (3T) data and applying them to unseen low-field MRCP (0.55T) acquisitions.

# Theory

## MRI Reconstruction

The MRI reconstruction model can be expressed as

$$y = \boldsymbol{E}x + n, \tag{1}$$

where $x$ is the final image to be recovered, $y$ the acquired multi-coil $k$-space data, and $n$ the noise. The linear encoding operator $\boldsymbol{E}$ includes coil sensitivity maps, the Fourier transform operator and the sampling pattern[20]. The MR image reconstruction problem becomes ill-posed problem in the case of undersampled acquistions and requires regularization. The optimization problem corresponding to a regularized reconstruction can be formulated as

$$\underset{x}{\operatorname{argmin}} \|y - \boldsymbol{E}x\|_2^2 + \lambda \mathcal{R}(x)\,, \tag{2}$$

where $\mathcal{R}$ is a regularization operator, and $\lambda > 0$ is a regularization parameter to balance the data fidelity and the regularization term. Classical choices of $\mathcal{R}$ include the $\ell_1$-norm of wavelet coefficients[12,21] or the Total Variation[22].

DL reconstruction can be motivated as an unrolled gradient decent algorithm of Equation ( 2 ) with a fixed number of iterations[23]. The neural network corresponds to the derivative of the regularizer $\mathcal{R}$ and a gradient descent step is given by

$$x^{i+1} = \underset{x}{\operatorname{argmin}}\ \|y - \boldsymbol{E}x\|_2^2 + \lambda \left\| x - \mathcal{N}_\theta\left(x^i\right) \right\|^2, \tag{3}$$

where $x^0$ is the initial reconstruction of the zero-filled $k$-space data and $\mathcal{N}_\theta$ the output of the neural network parameterized with $\theta$. The regularization parameter $\lambda$ is a trainable parameter along with $\theta$.

## Training strategies for Deep Learning-based MRI Reconstruction

### Supervised Learning

In a supervised training, the DL reconstruction model is trained with training data $\left\{\left(x_{ref}^j, y^j\right) : j = 0, \dots, M-1\right\}$, where $M$ denotes the number of training data, $x_{ref}^j$ and $y^j$ are the coil-combined ground truth images and the multi-coil undersampled $k$-space data at the $j$-th pair. $x_{ref}^j$ is a reconstruction from a fully sampled k-space that this is typically obtained with a fully-sampled acquisition. For our application of MRCP, we propose to use a GRAPPA[24] reconstructed k-space from an undersampled acquisition. The goal of supervised learning is to adjust the model parameters $\theta$ and the regularization parameter $\lambda$ in the DL reconstruction model from Equation ( 3 ) by minimizing the loss,

$$\underset{\theta,\lambda}{\operatorname{argmin}} \frac{1}{M} \sum_{j=0}^{M-1} \mathcal{L}\left(x_{ref}^{j}, f\left(y^{j}, \boldsymbol{E}^{j}; \theta, \lambda\right)\right),$$

( 4 )

where $f(y^j, \boldsymbol{E}^j; \theta, \lambda)$ denotes the output of the unrolled network according to the iterative algorithm in Equation ( 3 ) with a fixed number of iterations. $\mathcal{L}(\dots)$ measures the difference between ground truth and network output and is minimized over the full set of training data. An example for a commonly used loss function is the mean squared error[14].

### Self-supervised Learning

If the acquisition of fully-sampled k-space is not possible, the alternative to supervised learning from GRAPPA reconstructions is self-supervised learning. The SSDU approach (self-supervised learning via data undersampling)[15] splits the undersampled $k$-space data into two subsets for data fidelity and loss calculation as

$$\Omega = \Theta \cup \Lambda,$$

( 5 )

where $\Omega$ are the measured undersampled k-space data and $\Theta$ and $\Lambda$ are disjoint from each other. The loss function for self-supervised learning of the reconstruction that solves Equation **Error! Reference source not found.** is,

$$\underset{\theta,\lambda}{\operatorname{argmin}} \frac{1}{M} \sum_{j=0}^{M} \mathcal{L}\left(y_{\Lambda}^{j}, \boldsymbol{E}_{\Lambda}^{j}\left(f\left(y_{\Theta}^{j}, \boldsymbol{E}_{\Theta}^{j}; \theta, \lambda\right)\right)\right),$$

( 6 )

where $f\left(y_{\Theta}^{j}, \boldsymbol{E}_{\Theta}^{j}; \theta, \lambda\right)$ is the unrolled network output image using $k$-space data only at $\Theta$ indices. Then, the output image is transformed into $k$-space data using the encoding operator $\boldsymbol{E}_{\lambda}^{j}$, specified by the $\Lambda$ indices. Following the SSDU approach, we used a combination of the $\ell_1$-$\ell_2$ norm calculated between the subset $y_{\Lambda}^{j}$ and the output $k$-space at indices $\Lambda$ as the loss function in our experiments.

## Materials and Methods

### MRCP Data Acquisition

From February 2023 to February 2024, we collected MRCP data from 35 healthy volunteers (23 males and 12 females). All participants received an informed consent discussion and gave their written informed consent for their data being further used and processed. The mean age of the participants was 56.7, ranging from 20 to 81. Thirty-one out of the 35 volunteers were scanned at a field strength of 3T (MAGNETOM Vida and Lumina, Siemens Healthineers, Erlangen, Germany), and four volunteers at a field strength of 0.55T (MAGNETOM Free.Max, Siemens Healthineers, Erlangen, Germany). We employed multi-channel coil arrays for imaging, consisting of an 18-/12-/9-channel body array and a 36-

/24-/6-channel spine array, for the Vida (3T), Lumina (3T), and Free.Max (0.55T) scanners. The volunteers were positioned in the supine, head-first orientation during the scan. We acquired 3D MRCP data using a 3D T2-weighted turbo spin-echo sequence (SPACE)[25] in free-breathing, using the PACE triggering method for motion management. We averaged MR signals for 1.4 and 2 times for 3T and 0.55T scanner systems, respectively, to compensate the free induction decay (FID) artifacts, as commonly done for the SPACE sequence[26]. Furthermore, we accelerated MRCP acquisition by factors of R=2 and R=6 using regular equidistant undersampling along the phase encoding direction (fully-sampled in the read out and slice directions) with 24 autocalibration signal (ACS) lines for both acceleration rates. These ACS were integrated in the accelerated data during the reconstruction. Notably, we performed multiple scans of volunteers for the training dataset by varying the field-of-view. Table 1 shows detailed scan parameters.

## Raw Data Processing

We converted the raw data to the ISMRMRD format[27] using the ISMRMRD[*] Python toolbox. We converted 3D $k$-space data to a 2D format by applying an Inverse Fourier Transform (IFT) along the fully-sampled slice (partition) direction. We performed a volume-wise normalization on the stack of 2D $k$-space data to ensure consistent data scaling across volumes. We then undersampled the initial two-fold accelerated MRCP data by an additional factor of three to generate the six-fold undersampled input to train the reconstruction network.

We estimated coil sensitivity maps using the ESPIRiT algorithm[28] using the Sigpy Python package[29] with a $24 \times 24$ fully-sampled center $k$-space block and a $5 \times 5$ kernel. Since fully sampled 24 ACS were kept for all accelerations, the estimation of sensitivity maps was not impacted by the choice of R. Contrary to the default setting in Sigpy, we did not crop the background to zero in the image domain. We maintained consistency across all reconstructions in this study by estimating and using the same set of coil sensitivity maps for all reconstructions of each dataset.

## Deep Learning-based Reconstruction

### Dataset Preparation

We divided our group of 35 volunteers into three independent sets: 18 volunteers for training, 4 for validation, and 13 for test. The training datasets included 39 data volumes derived from multiple scans of the 18 volunteers, all obtained at 3T with six-fold retrospective undersampling. The validation dataset comprised 4 data volumes from 4 validation volunteers, also obtained at 3T with six-fold retrospective undersampling.

The test dataset consisted of 26 data volumes from 13 test volunteers, obtained with retrospective and prospective undersampling with R=6 from each volunteer. 9 out of 13 test volunteers were scanned at 3T, four at 0.55T.

We performed retrospective undersampling by omitting k-space lines from R=2 acquisitions to achieve R=6 undersampling along the phase-encoding direction while preserving the 24 autocalibration signal (ACS) lines. For prospective undersampling we directly acquired only the undersampled k-space lines during the scan.

[*] https://pypi.org/project/ismrmrd/

## Network Architectures and Training

We built a ResNet[30]-based DL reconstruction model for SV[14] and SSV[15]. We used the same DL model architecture for both trainings strategies, consisting of 12 unrolling steps and shared model weights over the steps. Each unrolling step consisted of eight residual blocks. We used the same pre-estimated coil sensitivity maps [28] to train both models.

For SV, we used zero-filled multi-coil $k$-space from retrospective undersampling of R=6 as the input and the GRAPPA reconstruction of R=2 as the reference for training. For SSV, the R=6 retrospectively undersampled data $\Omega$, were further split into two mutually exclusive subsets, $\Theta$ for data consistency and $\Lambda$ for loss calculation. Even though SSV can be trained directly using prospective R=6 undersampling, we used retrospective R=6 undersampling for a fair and consistent evaluation between SV and SSV. The subsets $\Theta$ and $\Lambda$ were sampled in a 1-D Gaussian distribution to match the aliasing distribution of $\Omega$[31]. We used the Adam optimizer with a learning rate of 0.0003 and minimized $\ell_1$- $\ell_2$ loss[15] in the $k$-space domain for both SV and SSV. We selected hyperparameters for the network models and trainings empirically.

## Conventional Reconstruction

We used iterative parallel imaging (CG-SENSE)[20] and $\ell_2$-Wavelet regularized compressed sensing (CS)[12] based on the Sigpy Python package[29] as reference reconstructions. Through empirical testing, we determined that the optimal regularization parameters for balancing SNR and aliasing artifacts were 0.01 for CG-SENSE and 0.008 for CS.

## Evaluation

We assessed the two DL models through quantitative and qualitative comparisons with CG-SENSE and CS. In the retrospective study, we computed average peak signal-to-noise ratio (PSNR) and structural similarity (SSIM) metrics for the test dataset. These metrics were computed on the magnitude-valued 3D reconstruction volume, specifically the stack of the absolute-valued 2D reconstructions. Reconstructions, including CG-SENSE, CS, SSV, and SV, on retrospective undersampling at R=6 were compared to the GRAPPA reconstruction at R=2. We also calculated the average PSNR and SSIM over 13 test datasets, nine for 3T and four for 0.55T. A one-sided Wilcoxon signed-rank test was performed to prove statistical significance of DL models against reconstructions in the comparison group in terms of PSNR and SSIM. A $p$-value of less than 0.05 was considered statistically significant. The null hypothesis stated that there is no statistically significant difference in the metric scores, while the alternative hypothesis posited that the DL models outperformed the other reconstructions. In the prospective study, we focused on perceptional image quality.

In addition, we obtained line profiles from reconstructions to analyze details of the hepatobiliary ducts and the gallbladder, as well as the presence of residual aliasing artifacts. The line profiles were manually selected based on the maximum intensity projection (MIP) image of the 3D reconstructed volume. Moreover, we calculated the Pearson product-moment correlation coefficient (PPMCC)[32] of the line profiles for all reconstructions with respect to the GRAPPA reconstruction at R=2. The PPMCC serves as a measure of similarity, with values closer to one indicating a higher degree of similarity between two profiles.

### Computational Resources

All experiments were conducted on a system equipped with an AMD Rome 7662 CPU 2.0 GHz, 512 GB of RAM, and an NVIDIA Tesla A100 SXM4 with 40 GB memory. Our model development was performed on a Linux Ubuntu v20.04.6 LTS environment, with Python v3.10.14, PyTorch v2.1.2, CUDA v11.8.0, PyTorch Lightning v2.2.5, and Sigpy v0.1.26.

## Results

### Retrospective study

Figure 1 depicts reconstructions for retrospective undersampling with R=6 at 3T and 0.55T, along with PSNR and SSIM metrics related to GRAPPA reconstruction at R=2. The top row of each subfigure presents MIP views. The red boxes on the MIP images determine the view of the cropped MIP images in the second row. The last row shows representative slices depicting differences in interesting anatomical details. Note that the metrics are calculated on 3D volumes, not on MIP images. In Figure 1(a), which shows results from the 3T scanners, CG-SENSE achieves PSNR (dB)/SSIM (%) values of 35.32/75.64, while CS, SSV, and SV show values of 38.83/82.46, 39.67/82.85, and 40.34/85.50. In addition, the blue circles in Figure 1(a) illustrate that DL reconstructions remove aliasing artifacts that are present in the conventional reconstructions. Furthermore, the orange circles in Figure 1(a) indicate that DL reconstructions preserve the sharpness of the common bile duct. In Figure 1(b), showing 0.55T results, CS-SENSE achieves PSNR (dB)/SSIM (%) values of 23.58/28.77, CS 30.38/59.20, SSV 31.71/69.93, and SV 32.17/71.41. The turquoise circles in Figure 1(b) demonstrate that DL reconstructions efficiently suppress background noise while preserving hepatobiliary ducts. Reconstructions of retrospective undersampling from additional volunteers can be found in Figures S1-S8 for 3T and S9-S11 for 0.55T.

Figure 2 contains violin plots of PSNR in dB and SSIM in % for the two employed field strengths. These metric scores are derived from nine volunteers at 3T and four volunteers at 0.55T in the test dataset. DL reconstructions consistently outperform conventional reconstructions. In particular, SV achieves the highest median values for both PSNR and SSIM metrics. SSV and CS achieve slightly lower metric scores. CG-SENSE consistently performs the worst across all experiments. Table 2 summarizes the average PSNR and SSIM of the test dataset, along with the corresponding $p$-values of the Wilcoxon signed-rank test at 3T and 0.55T. SV consistently achieves the highest metrics for both field strengths. Therefore, the other reconstruction methods are compared to SV for the Wilcoxon test. The Wilcoxon test results show statistical significance between SV and all other reconstructions at 3T, with all p-values below 0.05. At 0.55T, none of the comparisons yield statistically significant differences.

Figure 3 shows the line profiles at the end of the hepatobiliary bile ducts and the gallbladder. The positions of the line profiles, indicated by the red and green arrows in Figure 1(a), are consistently applied across all five reconstructions, ensuring accurate comparison. The line profiles of SV are very close to the ground truth (GRAPPA) and show the highest PPCMM scores of 0.988 and 0.999 for Figure 3(a) and Figure 3(b). The blue arrows in Figure 3(a) demonstrate that both SV and SSV effectively remove aliasing artifacts, which are present in the other reconstructions. In addition, the image intensity of the gallbladder is consistent between the DL reconstructions and the GRAPPA R=2 reconstruction (pink circle in Figure 3(b)).

## Prospective study

Table 1 includes the required breathing cycles and actual acquisition time (TA). For MRCP acquisitions at R=2, 3T and 0.55T require 97 and 82 breathing cycles. This corresponds to estimate times of 303 and 249 seconds based on the scanning protocol. In contrast, acquisitions at R=6 for both field strengths require only 39 and 38 cycles, with corresponding times of 138 and 139 seconds. A similar tendency is seen for TA, where the average TAs at R=2 are 599 and 542 seconds for 3T and 0.55T, and 255 and 180 seconds at R=6.

Figure 4 depicts the results of reconstructions for prospective undersampling with R=6 for 3T and 0.55T. Only DL reconstructions clearly remove aliasing artifacts, as shown in the blue circles of Figure 4(a). Fine hepatobiliary ducts, which are buried under the noise level in conventional reconstructions, become visible in the SV and SSV reconstructions (turquoise circles of Figure 4(b)). Results of prospective undersampling from additional volunteers are available in Figures S12-S19 for 3T and S20-S22 for 0.55T.

# Discussion

The goal of our study was to reduce the acquisition time for PACE-using higher acceleration factors with DL-based image reconstruction. We achieved accelerated PACE-MRCP acquisitions by undersampling three times more than in clinical applications. Specifically, the acquisition times at R=6 were 2.4 and 3.0 times faster than those at R=2 for 3T and 0.55T. DL reconstructions suppressed aliasing artifacts and noise amplification caused by undersampling with high acceleration factors. Our prospective experiments demonstrate that DL reconstructions can be applied to truly accelerated acquisitions on the scanner.

We chose a ResNet architecture[30] as the backbone for both our SV and SSV trainings to be consistent with the literature describing these two approaches[14,15]. However, it is worth noting that many DL reconstruction techniques use alternative architectures, such as the Variational Network (VN) [33]or its end-to-end variant[34] that utilizes a UNet. The deep learning models in our study can in principle be changed to different architectures with proper optimization and modifications to suit their specific characteristics but we cannot make any statements about their respective performances.

The Wilcoxon test reveals that there is no statistically significant difference between the reconstructions at 0.55T (Table 2). However, Figure 1 and Figure 4 demonstrate that SV consistently produces superior image quality in terms of aliasing artifacts, background noise level, and sharpness. This indicates that the metric scores for MRCP do not fully reflect the actual image quality, especially for the low field case. One potential explanation for this discrepancy is that all metric scores are calculated with GRAPPA R=2 acceleration. These reconstructions are not a true ground truth because they are already subject to some degree of noise amplification[35]. This effect of noise in the ground truth is particularly strong in MRCP because the majority of the individual slices are characterized by significant background noise. This makes the quantification of MRCP image quality with metrics like PSNR and SSIM challenging. Therefore, though metric scores can provide some quantitative guidance, it is important to prioritize visual image quality and avoid over-interpretation of these quantitative metric scores.

The limitation of quantitative metric scores is particularly relevant when comparing the performance of supervised trainings using R=2 GRAPPA reconstructions as the ground truth, and self-supervised trainings. While their image quality appears similar in Figures 1 and 4, supervised trainings consistently lead to higher quantitative metric scores. An explanation for this effect is that the goal of SV trainings was to minimize the difference between the output of the model and the GRAPPA R=2 reconstructions, which also served as the reference in the evaluation. In contrast, self-supervised trainings were performed directly on the undersampled data without the use of GRAPPA reconstructions.

Nevertheless, an interesting finding of our experiments is that, despite the ground truth for SV training not being fully sampled, SV performs as well as SSV. One possible explanation for this is that the acceleration factor used for our parallel imaging ground truth reconstructions followed the standard clinical MRCP protocol, which involved a relatively low value (R=2). Consequently, our GRAPPA reconstructions yielded clinically acceptable image quality without significant artifacts or noise amplification.

One open question in the field of deep learning MR image reconstruction is the generalization of a single model with respect to changes in the image acquisition setup. In our experimental design, we tested this by training our deep learning model on 3T data and then applying it to data acquired at 0.55T. Our results show that even with this domain shift between training and testing, the data-driven deep learning models consistently outperformed compressed sensing and parallel imaging, particularly in terms of noise suppression. This robustness was a surprising result to us because early work on the generalization potential of deep learning reconstructions showed that while trained models generalized well over multiple image contrasts, generalization was poor for changes in SNR[19].

There are some limitations in this study. First, we were unable to evaluate our proposed method using patient data or through radiologist interpretation studies. We acknowledge that real-world clinical data and expert evaluations are crucial for validating the practical applicability and diagnostic reliability of reconstruction techniques. Therefore, future studies should aim to incorporate patient datasets and assessments from radiologists to ensure that the method meets clinical standards and can be effectively used in practice.

The second limitation of our study is that it focuses only on 1D accelerations where k-space was always fully sampled along the slice direction, together with 2D slice-by-slice reconstructions. We chose this approach to simplify the study setup in light of the substantial GPU memory requirements for the training of deep learning reconstruction models for 3D MRCP data. As it is a general property of parallel imaging that improved image quality can be achieved with the introduction of multi-dimensional undersampling[36], we expect that all reconstruction methods tested in our study will show improved performance with 2D acceleration in combination with 3D reconstruction. Overcoming the GPU memory limitations for DL reconstruction is a goal of our future work.

## Conclusion

The results of our study demonstrate a significant reduction in acquisition time for PACE-MRCP with DL reconstruction. DL resulted in superior performance in comparison to conventional parallel imaging and compressed sensing reconstructions for both retrospectively and prospectively accelerated acquisitions. Our results also demonstrated generalizability of MRCP DL reconstructions across field strengths from 3T to 0.55T.

## Acknowledgements

This work is supported by the Erlangen National High Performance Computing Center (NHR@FAU) of Friedrich-Alexander-Universität Erlangen-Nürnberg(FAU) under the NHR project b143dc. NHR funding is provided by federal and Bavarian state authorities. NHR@FAU hardware is partially funded by the German Research Foundation (DFG) – 440719683. The authors thank Thomas Benkert, Bruno Riemenschneider, Zhengguo Tan, and Marc Vornehm for the beneficial discussion and technical supports.

## Conflict of Interest Statement

J.K. receives a PhD stipend from Siemens Healthineers AG. M.D.N. is employed by Siemens Healthineers AG. F.K. receives patent royalties for deep learning image reconstruction and research support from Siemens Healthineers AG, has stock options from Subtle Medical and is a consultant for Imaginostics.

## Data Availability Statement

The data that support the findings of this study are openly available in MRCP_DLRecon at https://doi.org/10.5281/zenodo.13912092.

## Funding Information

DFG, Grant/Award Number: 513220538; NIH, Grant/Award Number: R01EB024532, P41EB017183.

## Correspondence

Jinho Kim, Computational Imaging Lab, Department of Artificial Intelligence in Biomedical Engineering, Friedrich-Alexander-Universität Erlangen-Nürnberg, Werner-Von-Siemens-Straße 61, D-91054, Erlangen, Germany. Email: jinho.kim@fau.de

## Abbreviations

MRCP, magnetic resonance cholangiopancreatography; DL, deep learning; SV, supervised training; SSV, self-supervised training; SSDU, self-supervised learning via data undersampling; PI, parallel imaging; CG-SENSE, conjugate gradient sensitivity encoding imaging; CS, compressed sensing imaging; FID, free induction decay; IFT, inverse Fourier transform; PACE, prospective acquisition correction technique; MIP, maximum intensity projection; PPMCC, Pearson product-moment correlation coefficient, PSNR, peak signal-to-noise ratio; SSIM, structural similarity; SNR, signal-to-noise ratio; ACS, autocalibration signal;

## Supporting Information

Additional supporting information can be found online in the Supporting Information section.

# Figures

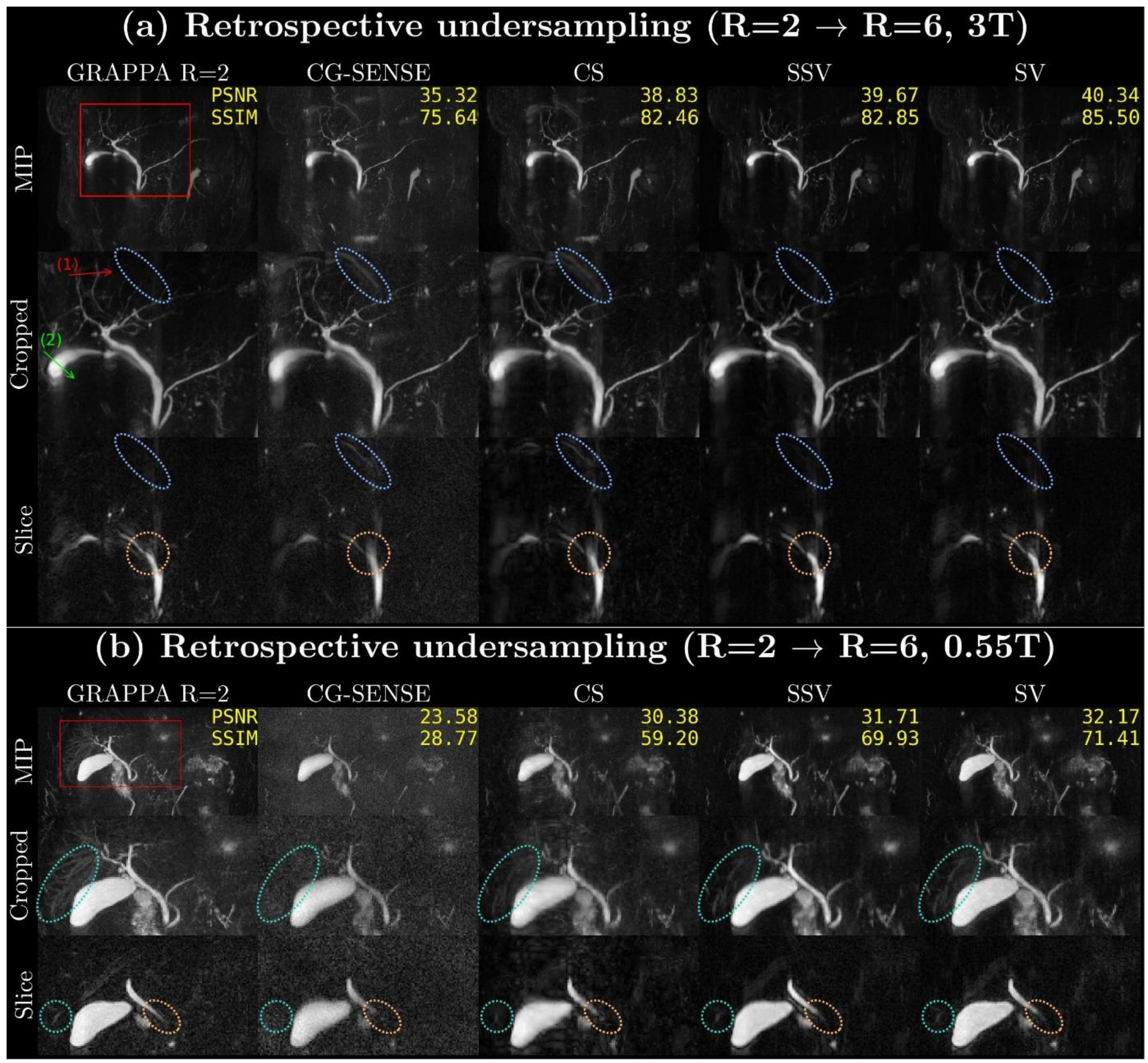

*Figure 1: Results of retrospective undersampling at (a) 3T and (b) 0.55T: We use GRAPPA with two-fold acceleration as the ground truth for comparison with CG-SENSE, CS, SSV, and SV reconstructions with six-fold acceleration. Each column corresponds to a reconstruction method and shows three different presentation forms: Maximum Intensity Projection (MIP) (top), a cropped view of the hepatobiliary duct of the MIP (middle), and a representative single slice (bottom). In (a), the blue circles indicate aliasing artifacts, and the orange circles display the representative sharpness of the common bile duct. In (b), the turquoise circles show details of the hepatobiliary ducts. PSNR in dB and SSIM in % of the six-fold reconstructions with respect to GRAPPA R=2 are shown in the top-right corner of the MIPs. The red and green arrows in (a) indicate the signal intensity source for the line profile representation for Figure 3.*

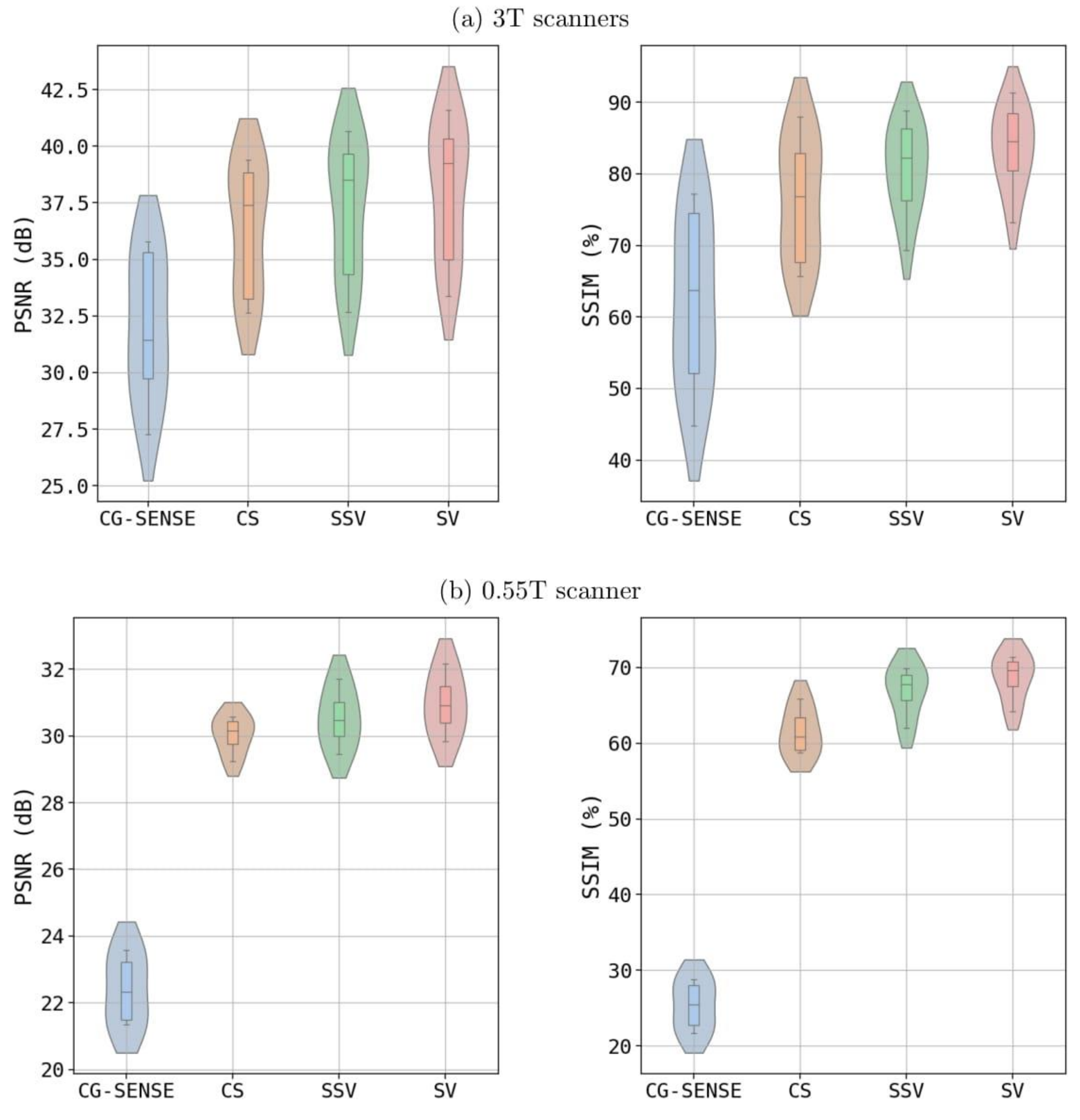

*Figure 2: Violin plots of PSNR in dB and SSIM in % of the different reconstruction methods against the R=2 GRAPPA reconstructions for (a) 3T (nine test set subjects) and (b) 0.55T (four test set subjects). Within each graph, box plots are included to provide a visual representation of the data distribution range.*

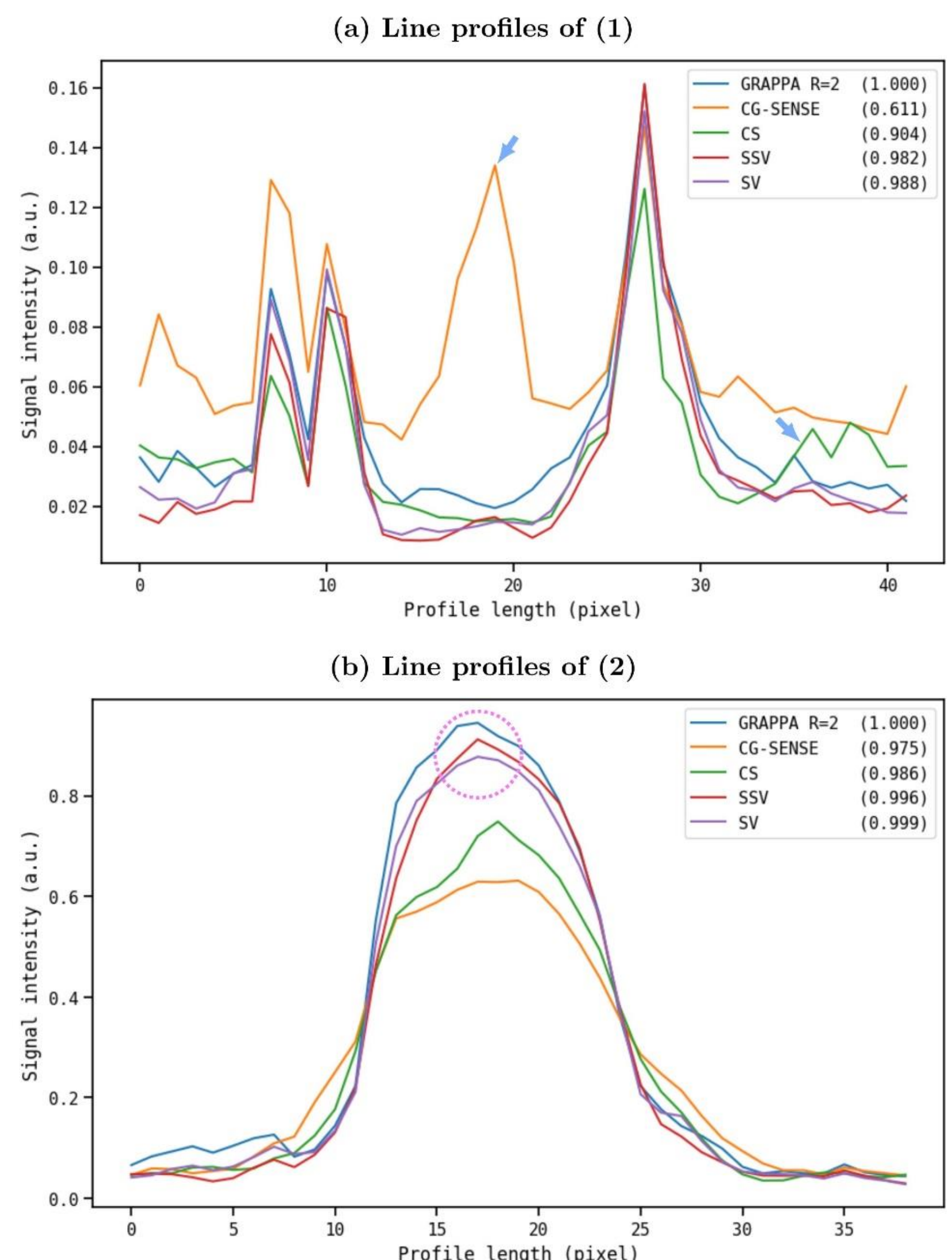

*Figure 3: Line profiles from the retrospective study. The line profiles are taken from the same position of the corresponding reconstructions and one representative position is presented as the red arrow (a) and the green arrow (b) in Figure 1(a). Pearson product-moment correlation coefficients (PPMCC) against GRAPPA R=2 are shown in the legend for the corresponding reconstructions to quantify the similarity to the reference profile. SV presents the highest PPMCC among others, meaning that the line profile of SV is the most similar to GRAPPA R=2. The blue arrows in (a) indicate aliasing artifacts of CG-SENSE and CG. The pink circle in (b) demonstrates that DL reconstructions preserve the signal intensity of the gallbladder.*

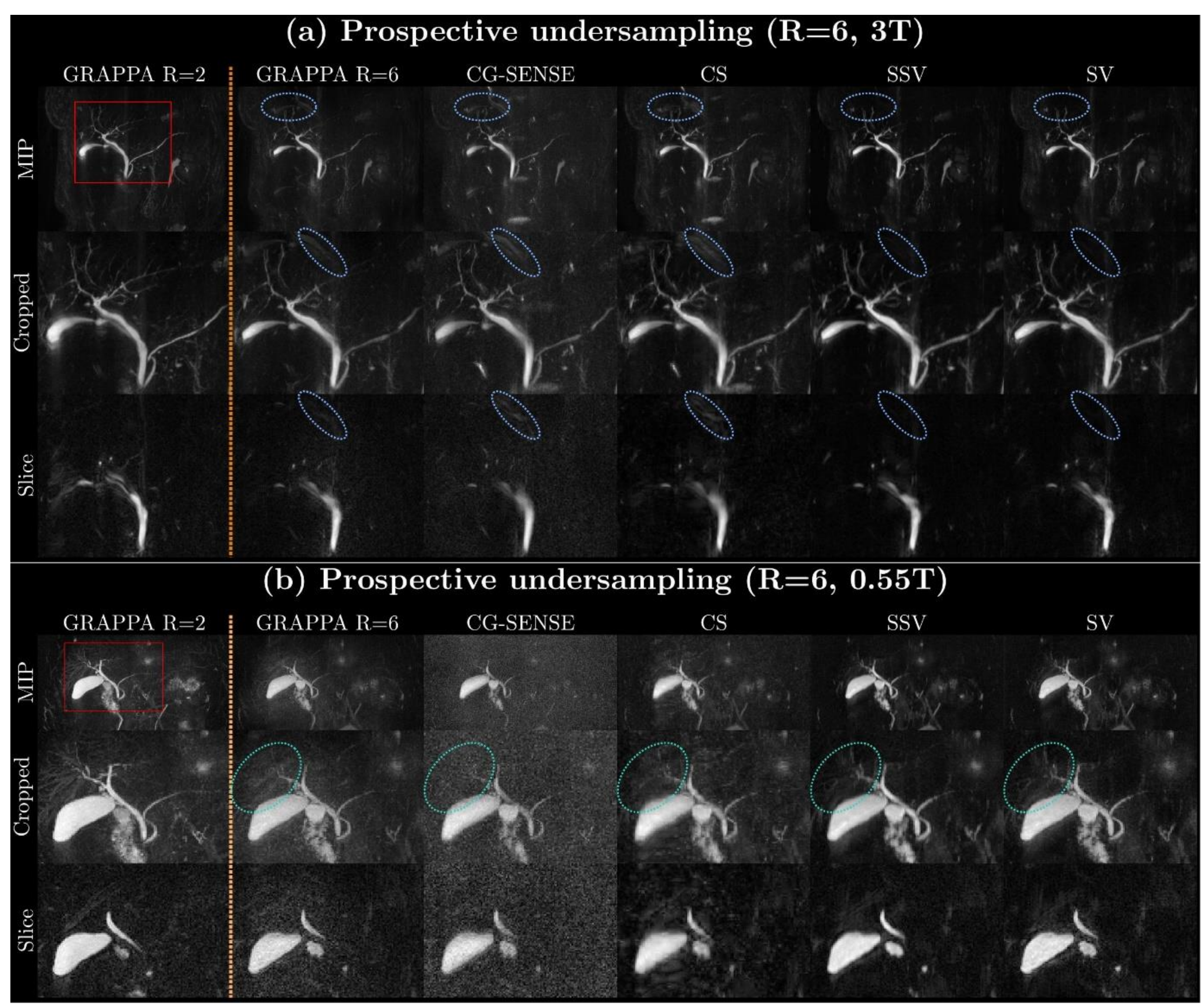

*Figure 4: Results of prospective undersampling at (a) 3T and (b) 0.55T: We use GRAPPA with a two-fold acceleration of the same volunteer as a visual reference (GRAPPA R=2). We reconstructed six-fold accelerated acquisitions using GRAPPA, CG-SENSE, CS, SSV, and SV. Each column corresponds to one reconstruction method and shows three different presentation forms: Maximum Intensity Projection (MIP) (top), a cropped view of the hepatobiliary duct of the MIP (middle), and a representative single slice (bottom). In (a), the blue circles indicate aliasing artifacts. In (b), the turquoise circles highlight details of the hepatobiliary ducts.*

# Tables

*Table 1: Parameters for MRI protocols*

| **Parameters** | **3T** | | **0.55T** | |
|---|---|---|---|---|
| Scanner | MAGNETOM Vida & Lumina | | MAGNETOM Free.Max | |
| Sequence | 3D T2-weighted TSE (SPACE) [20] | | | |
| Acquisition plane | Coronal | | | |
| Turbo factor | 180 | | | |
| TR (ms)* | 5985 ± 2103 (2700–12829) | | 5298 ± 1462 (4251–8878) | |
| TE (ms)* | 703 ± 2 (701–709) | | 703 | |
| Acquired voxel size (mm$^3$) | 0.5 × 0.5 × 1.2 | | 0.7 × 0.7 × 1.0 | |
| Matrix | 384 × 480 | | 256 × 194 | |
| Number of slices | 64 | | | |
| Flip angles* | 115, 120 | | 145 | |
| Echo train spacing (ms) | 5.26 | | 5.58 | |
| Echo train duration (ms) | 957 | | 1016 | |
| Number of signal averages | 1.4 | | 2 | |
| Bandwidth (Hz/pixel) | 352 | | 391 | |
| Triggering | PACE signal | | | |
| Number of ACS lines | 24 | | | |
| Parallel imaging acceleration PE | x2 | x6 | x2 | x6 |
| Parallel imaging acceleration 3D | x1 | | | |
| Breathing cycles (estim. time (s)) | 97 (303) | 39 (138) | 82 (249) | 38 (139) |
| TA (s)* | 599 ± 283 (229–1386) | 255 ± 81 (161–426) | 542 ± 172 (370–720) | 180 ± 28 (158–226) |

TSE, turbo spin-echo; TR, repetition time; TE, echo time; ACS, autocalibration signal; PE, phase encoding; TA, acquisition time; estim. time, estimated time.
The notation format for TR, TE, and TA is Mean ± Standard deviation (Minimum – Maximum).
* Variable depending on the volunteer.

*Table 2: Quantitative analysis of the retrospective study at R=6*

| | | **CG-SENSE** | **CS** | **SSV** | **SV** | ***p*-value** | | |
|---|---|---|---|---|---|---|---|---|
| | | | | | | CG-SENSE vs. SV | CS vs. SV | SSV vs. SV |
| 3T | PSNR | 32.05 ± 3.01 | 36.44 ± 2.69 | 37.29 ± 2.79 | **38.02 ± 2.81** | 0.002* | 0.002* | 0.002* |
| | SSIM | 62.08 ± 11.22 | 76.52 ± 8.06 | 80.79 ± 5.88 | **83.63 ± 5.38** | 0.002* | 0.006* | 0.002* |
| 0.55T | PSNR | 22.39 ± 0.97 | 30.03 ± 0.51 | 30.53 ± 0.82 | **30.96 ± 0.86** | 0.062 | 0.062 | 0.062 |
| | SSIM | 25.34 ± 3.0 | 61.58 ± 2.85 | 66.9 ± 3.01 | **68.72 ± 2.77** | 0.062 | 0.062 | 0.062 |

CG-SENSE, conjugate gradient SENSE; CS, compressed sensing; SSV, self-supervised model; VN, supervised model; PSNR, peak signal-to-noise ratio; SSIM, structural similarity.
The notation format for PSNR and SSIM is Mean ± Standard deviation.
PSNR and SSIM are averaged over nine test data for 3T and four for 0.55T.
* $p$-value $< 0.05$ is considered statistically significant.